\documentclass[showpacs, oneside, twocolumn, prl, amsmath, amssymb, nofootinbib, superscriptaddress]{revtex4-1}

\usepackage{cases}
\usepackage{amsmath}
\usepackage{amssymb}
\usepackage{amsfonts}
\usepackage{amssymb}
\usepackage{dcolumn}
\usepackage{bm}
\usepackage{bbm}
\usepackage{graphicx}
\usepackage{xcolor}
\usepackage{array}
\usepackage{subfigure}
\usepackage{hyperref}
\usepackage{wasysym}

\newcommand{\be}{\begin{equation}}
\newcommand{\ee}{\end{equation}}
\newcommand{\ba}{\begin{eqnarray}}
\newcommand{\ea}{\end{eqnarray}}

\newcommand{\gsim}{\mathrel{\hbox{\rlap{\lower.55ex \hbox {$\sim$}}
                   \kern-.3em \raise.4ex \hbox{$>$}}}}
\newcommand{\lsim}{\mathrel{\hbox{\rlap{\lower.55ex \hbox {$\sim$}}
                   \kern-.3em \raise.4ex \hbox{$<$}}}}

\hypersetup{colorlinks=true,
            breaklinks=true,
            pdfstartview=Fit,
            linkcolor=blue,
            citecolor=blue,
            urlcolor=blue}

\begin{document}
\title{Primordial Black Holes from Sound Speed Resonance during Inflation}
%\title{Primordial Black Holes from the Sound Speed Resonance Effect} %in Inflationary Cosmology}

\author{Yi-Fu Cai}
\email{yifucai@ustc.edu.cn}
\affiliation{CAS Key Laboratory for Researches in Galaxies and Cosmology, Department of Astronomy, University of Science and Technology of China, Hefei, Anhui 230026, China}
\affiliation{School of Astronomy and Space Science, University of Science and Technology of China, Hefei, Anhui 230026, China}

\author{Xi Tong}
\email{tx123@mail.ustc.edu.cn}
\affiliation{CAS Key Laboratory for Researches in Galaxies and Cosmology, Department of Astronomy, University of Science and Technology of China, Hefei, Anhui 230026, China}
\affiliation{School of Astronomy and Space Science, University of Science and Technology of China, Hefei, Anhui 230026, China}
\affiliation{Department of Physics, The Hong Kong University of Science and Technology, Clear Water Bay, Kowloon, Hong Kong, China}

\author{Dong-Gang Wang}
\email{wdgang@strw.leidenuniv.nl}
\affiliation{Lorentz Institute for Theoretical Physics, Leiden University, 2333 CA Leiden, The Netherlands}
\affiliation{Leiden Observatory, Leiden University, 2300 RA Leiden, The Netherlands}

\author{Sheng-Feng Yan}
\email{sfyan22@mail.ustc.edu.cn}
\affiliation{CAS Key Laboratory for Researches in Galaxies and Cosmology, Department of Astronomy, University of Science and Technology of China, Hefei, Anhui 230026, China}
\affiliation{School of Astronomy and Space Science, University of Science and Technology of China, Hefei, Anhui 230026, China}

\begin{abstract}

We report on a novel phenomenon of the resonance effect of primordial density perturbations arisen from a sound speed parameter with an oscillatory behavior, which can generically lead to the formation of primordial black holes in the early Universe. For a general inflaton field, it can seed primordial density fluctuations and their propagation is governed by a parameter of sound speed square. Once if this parameter achieves an oscillatory feature for a while during inflation, a significant non-perturbative resonance effect on the inflaton field fluctuations takes place around a critical length scale, which results in significant peaks in the primordial power spectrum. By virtue of this robust mechanism, primordial black holes with specific mass function can be produced with a sufficient abundance for dark matter in sizable parameter ranges.
\end{abstract}


\maketitle

%\section{Introduction}
{\it Introduction.} --
Investigations on primordial black holes (PBHs) offer an inspiring possibility to probe physics in the early Universe \cite{Zeldovich:1966, Hawking:1971ei, Carr:1974nx}. In recent years, the cosmological implications of PBHs have been extensively studied, especially since they could be a potential candidate for dark matter (DM) \cite{Ivanov:1994pa, Carr:2016drx, Gaggero:2016dpq, Inomata:2017okj, Georg:2017mqk, Fuller:2017uyd, Kovetz:2017rvv}. Moreover, the PBHs can also be responsible for some gravitational wave (GW) events \cite{Bird:2016dcv, Clesse:2016vqa, Sasaki:2016jop, Nakamura:1997sm}, for instance, the first direct detection of the GW event announced by the LIGO collaboration \cite{Abbott:2016blz}. In the literature, many theoretical mechanisms producing PBHs rely on a spectrum of primordial density fluctuations with extra enhancement on certain length scales, which are usually accomplished by a particularly tuned background dynamics of the quantum fields in the early Universe (e.g. see \cite{GarciaBellido:1996qt, Garcia-Bellido:2017mdw, Domcke:2017fix, Kannike:2017bxn, Carr:2017edp, Ballesteros:2017fsr, Hertzberg:2017dkh, Franciolini:2018vbk, Kohri:2018qtx, Ozsoy:2018flq, Biagetti:2018pjj} for various analyses within inflationary cosmology, see \cite{Chen:2016kjx, Quintin:2016qro} for the investigations within bounce cosmology, and see \cite{Khlopov:2008qy, Sasaki:2018dmp} for comprehensive reviews).

Primordial density fluctuations, that seeded the large-scale structure (LSS) of the Universe, are usually thought to arise from quantum fluctuations during a dramatic phase of expansion at early times, as described by inflationary cosmology, from which  a nearly scale-invariant power spectrum with a standard dispersion relation is obtained \cite{Mukhanov:1990me}. This was confirmed by various cosmological measurements such as the cosmic microwave background (CMB) radiation and LSS surveys at extremely high precision. It is interesting to note that, however, as advocated by theoretical developments of quantum gravity, modifications of the dispersion relation of primordial density fluctuations are naturally expected \cite{ArmendarizPicon:2003ht, ArmendarizPicon:2006if, Magueijo:2008sx, Cai:2009hc}, which could have non-trivial phenomenological consequences, as we will illustrate in this {\it Letter}.

%\section{Formalisms}
{\it Sound speed resonance.} --
We begin with a general discussion on the dynamics of primordial cosmological perturbations in the standard inflationary paradigm. The causal mechanism for generating a nearly scale-invariant primordial power spectrum suggests that, cosmological fluctuations should initially emerge inside a Hubble radius, and then leave it in the primordial epoch, and finally re-enter at late times. One uses a gauge-invariant variable $\zeta$, the curvature perturbation in comoving gauge, to characterize primordial inhomogeneities. For the general case with a non-trivial sound speed $c_s$ \cite{ArmendarizPicon:1999rj, Garriga:1999vw}, one can use a canonical variable $v \equiv z\zeta$, where $z\equiv \sqrt{2\epsilon}a/c_s$ with $\epsilon\equiv -\dot H/H^2$. The perturbation equation for a Fourier mode $v_k(\tau)$ in the context of General Relativity is given by:
$v_k'' + \big( c_s^2 k^2 - {z''}/{z} \big) v_k=0$,
where the prime denotes the derivative w.r.t. the conformal time $\tau$.

To generate PBHs within inflationary cosmology, one needs to consider how to amplify curvature perturbations for certain ranges of modes. In the literature, most studies focuses on non-conventional behaviours of inflationary background, such as a sudden change of the slow-roll parameter $\epsilon$ in the inflection-point inflation \cite{Garcia-Bellido:2017mdw}. Soon it was realized that, a fine-tuning of parameters is inevitable in this mechanism to obtain a sufficient enhancement of $\zeta$ to generate PBHs in abundance \cite{Germani:2017bcs, Motohashi:2017kbs}. In this {\it Letter}, we explore a novel possibility -- a parametric amplification of curvature perturbations caused by resonance with oscillations in the sound speed of their propagation, which, as we will show, provides a much more efficient way to enhance the primordial power spectrum around the astrophysical scales where PBHs could account for DM in the current experimental bounds.

The sound speed parameter $c_s$ can deviate from unity during the primordial era, namely, in a general single-field model with a non-canonical kinetic term. This arises when inflation models are embodied in UV-complete theories, such as D-brane dynamics in string theory \cite{Silverstein:2003hf, Alishahiha:2004eh}, or, from the effective field theory viewpoint, when heavy modes are integrated out \cite{Achucarro:2010da, Achucarro:2012sm}. How variation of the primordial sound speed affects curvature perturbations has already been extensively studied, but mainly in the context of primordial features on CMB scales \cite{Achucarro:2013cva, Chluba:2015bqa}.

In this {\it Letter}, we put aside the theoretical constructions, and take a phenomenological approach to study the effects of an oscillating sound speed on the power spectrum at much smaller scales. As a starting point, we consider the following parametrization: 
\begin{eqnarray}\label{eq:param_cs}
 c_s^2 = 1 -2 \xi \big[ 1-\cos(2k_*\tau) \big] ~, ~~~~\tau>\tau_0~,
\end{eqnarray}
where $\xi$ is the amplitude of the oscillation and $k_*$ is the oscillation frequency. Note that, $\xi < 1/4$ is required such that $c_s$ is positively definite. The oscillation begins at $\tau_0$, where $k_*$ is deep inside the Hubble radius, {\it i.e.} $|k_*\tau_0|\gg1$. To simplify the analysis we set $c_s=1$ before $\tau_0$ and let it transit to the oscillatory regime smoothly. To demonstrate how sound speed oscillations affect perturbations, we approximately treat inflation as a de Sitter (dS) expansion \footnote{Here the dS approximation is good enough to discuss the effects of the sound speed resonance, as we will justify later with numerical analysis.}. By applying the parametrization in Eq.~\eqref{eq:param_cs}, the effective mass term in the perturbation equation can be expanded as:
\begin{eqnarray} \label{effmass}
 \frac{z''}{z}=\frac{2}{\tau^2}-\frac{4\xi k_*}{\tau}\sin(2k_*\tau)+4\xi k_*^2 \cos(2k_*\tau)+\mathcal{O}(\xi^2)~. 
\end{eqnarray}
As we are interested in the modes on sub-Hubble scales, the first two terms become negligible and the perturbation equation can be approximately written as:
\begin{eqnarray}\label{eom2}
 \frac{d^2v_k}{dx^2} + \big( A_k - 2q \cos 2x \big) v_k =0 ~,
\end{eqnarray}
where $x\equiv-k_*\tau$, $A_k={k^2}/{k_*^2}+2q-4\xi$ and $q=2\xi-({k^2}/{k_*^2})\xi$. This is the Mathieu equation, which presents a parametric instability for certain ranges of $k$. This equation has been widely applied in the preheating stage after inflation, where excitations of an additional particle can be resonantly amplified, leading to an efficient energy transfer from the inflaton into other particles (see \cite{Traschen:1990sw, Kofman:1994rk, Kofman:1997yn} for early studies and see \cite{Bassett:2005xm, Allahverdi:2010xz} for comprehensive reviews). For the process of preheating, the parametric resonance of fluctuations is driven by oscillations of the inflaton field, leaving the possibility of an amplification of the perturbation modes in the whole infrared regime.

\begin{figure}[h!]
\centering
\includegraphics[width=0.45\textwidth]{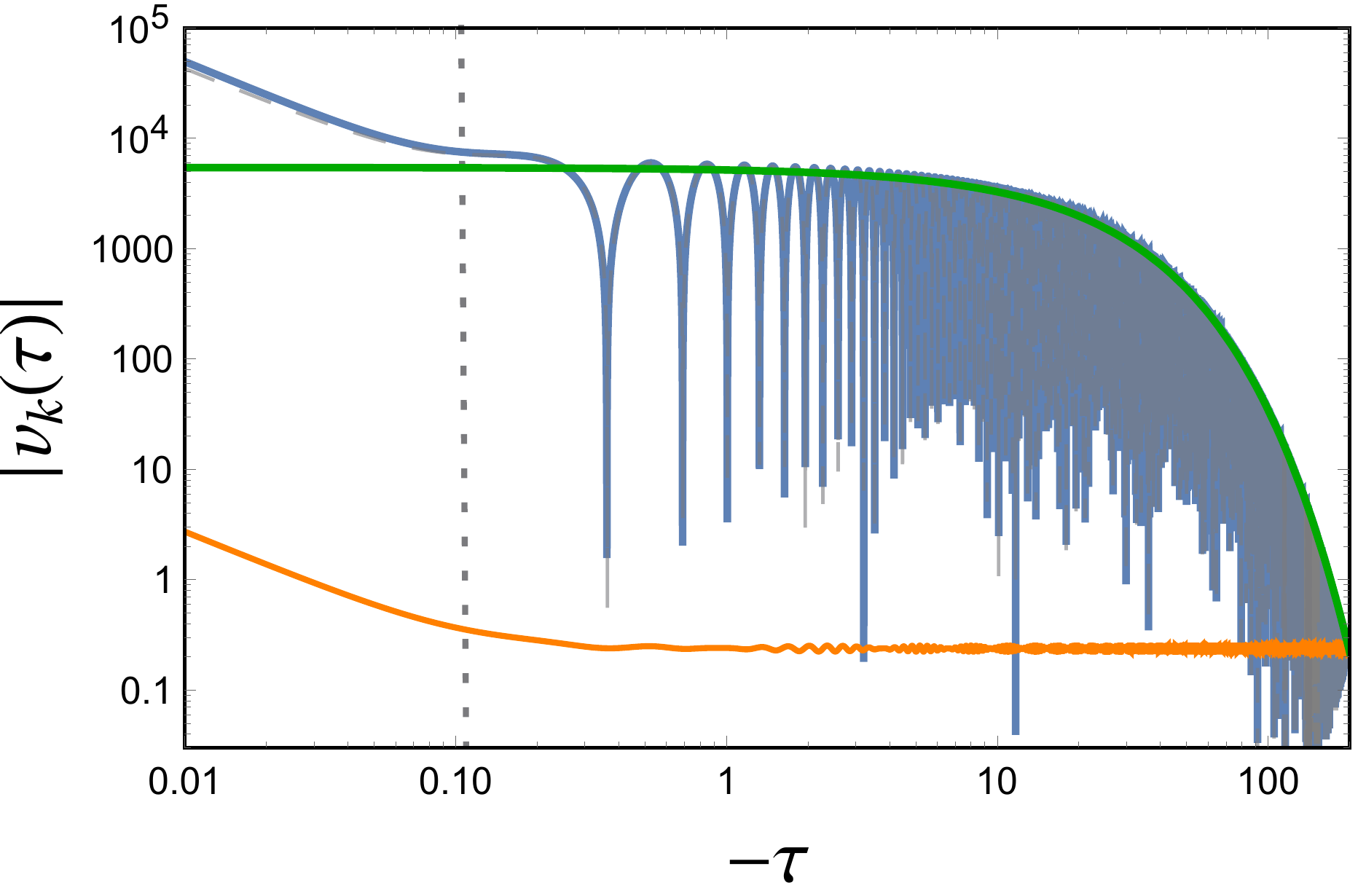}
\caption{
Parametric amplification of the resonating $k_*$ mode. The conformal time evolves from right to left. Numerical solutions under dS approximation and in the Starobinsky inflation model are given by the blue solid curve and the grey dashed line, respectively. The green line is the analytical profile of Eq.~\eqref{profile}, and the orange line represents a mode $k \neq k_*$ that is not resonating. The vertical dotted line depicts the time of Hubble-crossing for the $k_*$ mode. 
}
\label{fig1:resonance}
\end{figure}

In our case the parametric resonance is seeded by an oscillatory contribution in the sound speed during inflation. In addition, as $\xi$ is always small and thus $|q|\ll1$, resonance bands are located in narrow ranges around harmonic frequencies $k\simeq nk_*$ of the oscillating sound speed. Since the first band ($n=1$) is significantly more enhanced than the subsequent harmonic bands, in the following analyses we focus on the resonated modes around $k_*$. With dS approximation, we numerically solve the perturbation equation with effective mass given in Eq.~\eqref{effmass}. By setting the mode function at the beginning of resonance to the Bunch-Davies vacuum $v_k(\tau_0)=e^{-ik\tau_0}/\sqrt{2k}$, we get the approximated numerical solution of $v_k(\tau)$ as {depicted by the blue solid curve} in Fig.~\ref{fig1:resonance}. We see that, for modes $k\neq k_*$ that are not resonating, $v_k(\tau)\sim const.$ inside the Hubble radius, and $\sim 1/\tau$ after Hubble-exit: they evolve as usual in the Bunch-Davies state. Meanwhile, the $k_*$ mode enters in resonance. On sub-Hubble scales, its exponential growth can be captured by:
\be \label{profile}
 v_k(\tau)\propto\exp(\xi k_*\tau/2)~,
\ee
as shown by the green curve in Fig.~\ref{fig1:resonance}. The amplification ceases around the Hubble-exit, since the first term in Eq.~\eqref{effmass} becomes dominant on super-Hubble scales. To check the validity of dS approximation, we also present the numerical result by considering a specific inflation model as an example, namely, the Starobinsky potential \cite{Starobinsky:1980te}, of which the background evolution is a quasi-dS expansion. The numerical solution of $v_k(\tau)$ in this model is shown by the grey dashed line in Fig. \ref{fig1:resonance} and matches the approximated one by the blue solid curve very well.

\begin{figure}[h!]
\centering
\includegraphics[width=0.4\textwidth]{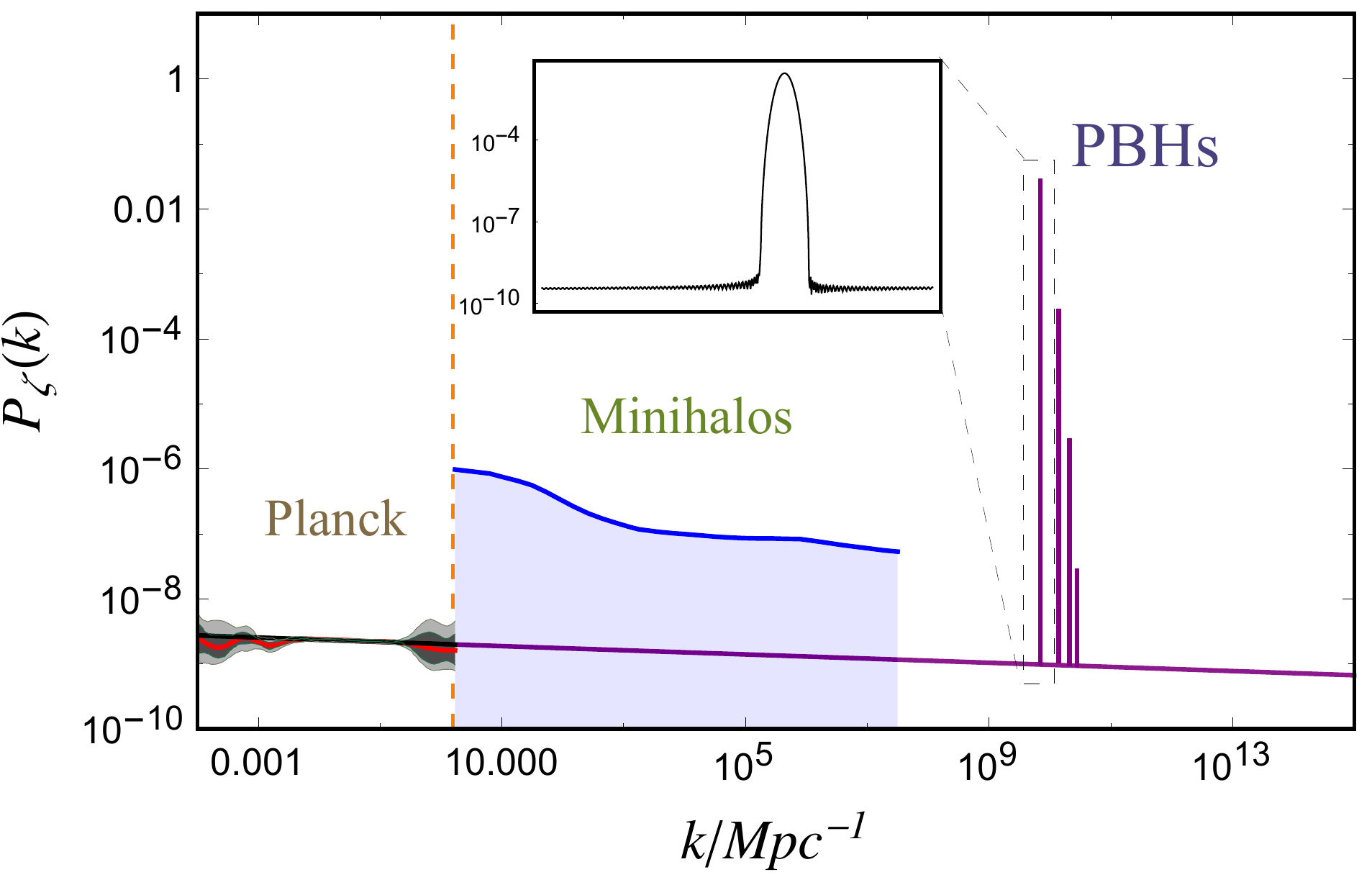}
\caption{
The power spectrum of primordial curvature perturbations with sharp peaks caused by sound speed resonance, and the comparison with various observational windows \cite{Bringmann:2011ut}. The first peak around the resonating mode $k_*=7\times10^{9} ~\text{Mpc}^{-1}$  is the most significant one, while others at subsequent harmonics $2k_*$, $3k_*$, $4k_*$ $...$ are sub-dominant by at least two orders of magnitude.
}
\label{fig2:spectr}
\end{figure}

For curvature perturbation, the $k_*$ mode evolves as $\zeta_{k_*}(\tau) \simeq \zeta_{k_*}(\tau_0) e^{\xi k_* (\tau-\tau_0)/2} {\tau}/{\tau_0}$ before Hubble-crossing.
It freezes at Hubble-exit, $\tau_*=-1/k_*$, with an enhanced amplitude:
\be
\zeta_{k_*} \simeq \zeta_{k_*}(\tau_0)
  \left(\frac{-1}{k_*\tau_0}\right)e^{-\xi k_*\tau_0/2} \simeq
  \frac{H}{\sqrt{4\epsilon k_*^3}}e^{-\xi k_*\tau_0/2}~, \nonumber
\ee
where in the second equality we have used $\zeta_{k_*}(\tau_0)=\frac{-H\tau_0}{\sqrt{4k_*\epsilon}}$, as given by the Bunch-Davies vacuum. The resulting primordial power spectrum $P_\zeta\equiv k^3|\zeta_k|^2/(2\pi^2)$ presents the following feature: while for modes $k \neq k_*$, we get the standard scale-invariant result, $P_\zeta=\frac{H^2}{8\pi^2\epsilon}$, there is a significant peak from the exponential amplification at the resonance frequency $k_*$, $P_\zeta=\frac{H^2}{8\pi^2\epsilon}e^{-\xi k_*\tau_0}$, as shown in Fig.~\ref{fig2:spectr}. The enhancement factor $e^{-\xi k_*\tau_0}$ arises from the interplay of two effects: the oscillation in the sound speed, controlled by its amplitude $\xi$, and the expansion of the Universe from the beginning of the resonance to Hubble-crossing of the $k_*$-mode: $-k_*\tau_0=\tau_0/\tau_*\simeq e^{\Delta N}$, where $\Delta N$ is the e-folding number for this period of inflation. By a rough estimate, even for very small oscillation amplitudes  $\xi\sim 10^{-4}$, a few e-folds $\Delta N \simeq 12$ is enough to get a peak of order unity in the power spectrum. Fig.~\ref{fig2:spectr} also shows peaks for harmonic frequencies $2k_*$, $3k_*$, $4k_*$, $...$, with relatively much lower amplitudes. For simplicity, we keep the discussion only on the $k_*$ mode, and parametrize the power spectrum using a $\delta$-function:
\begin{equation} \label{powerspec}
 P_\zeta(k)\simeq A_s \Big( \frac{k}{k_p} \Big)^{n_s-1} \Big[ 1+\frac{\xi k_*}{2}e^{-\xi k_*\tau_0} \delta(k-k_*) \Big]~,
\end{equation}
where $A_s = \frac{H^2}{8\pi^2\epsilon}$ is the amplitude of power spectrum as in standard inflation and $n_s$ is the spectral index at pivot scale $k_p\simeq 0.05 ~ \text{Mpc}^{-1}$ \cite{Ade:2015lrj}. The coefficient in front of the $\delta$-function is determined by estimating the area of the peak using a triangle approximation.

%\section{Results}
{\it The PBH formation.} --
We now study the formation of PBHs due to the enhancement in the primordial power spectrum. As we see, the width of the peak in the power spectrum being very narrow ($\sim \xi k_*$), only modes very close to the resonance frequency $k_*$ may have sufficiently large amplitude to collapse into black holes. After Hubble-exit, if the density perturbations from these modes are larger than a critical value $\delta_c$, then, after Hubble re-entry, they could collapse into black holes due to gravitational attraction. The Schwarzschild radius of PBHs with mass $M$ is related to the physical wavelength of the mode $k_M$ at Hubble re-entry, 
$k_{M,\mathrm{ph}} = k_M/a_M \simeq R_S^{-1} = \big( \frac{M}{4\pi M_p^2} \big)^{-1}$.
Accordingly, the PBH mass can be expressed as a function of $k_M$ via:
\begin{equation}\label{MPBH}
 M\simeq \gamma \frac{4\pi M_p^2}{H(t_\text{exit}(k_M))}e^{\Delta N(k_M)} ~,
\end{equation}
where $\Delta N(k_M)=\ln[a(t_\text{re-entry}(k_M))/a(t_\text{exit}(k_M))]$ is the the e-folding number from the Hubble-exit time of the mode $k_M$ to its re-entry. The correction factor $\gamma$ represents the fraction of the horizon mass responsible for PBH formation, which can be simply taken as $\gamma\simeq 0.2$ \cite{Carr:1975qj}. Given the sharpness of the peak in the power spectrum, the PBHs formed in this context are likely to possess a rather narrow range of masses.

To estimate the abundance of PBHs with mass $M$, one usually defines $\beta(M)$ as the mass fraction of PBHs against the total energy density at the formation, which can be expressed as an integration of the Gaussian distribution of the perturbations:
\begin{align}
 \beta(M) \equiv \frac{\rho_{\rm PBH}(M)}{\rho_{\rm tot}} = \frac{\gamma}{2} \mathrm{Erfc}[\frac{\delta_c}{\sqrt{2}\sigma_M}] ~,
\end{align}
where
$\mathrm{Erfc}$ denotes the complementary error function. $\sigma_M$ is the standard deviation of the density perturbations at the scale associated to the PBH mass $M$, which can be expressed as
$\sigma_M^2 = \int_0^\infty\frac{dk}{k}W(k/k_M)^2\frac{16}{81}\big(\frac{k}{k_M}\big)^4P_\zeta(k)$,
where $W(x)=\exp(-x^2/2)$ is a Gaussian window function. Since the scale-invariant part of the power spectrum is smaller than the critical density, no black holes would form except at scales around the resonance peak. As we are working in the perturbative regime, the height of the peak in $\sqrt{P_\zeta(k)}$ should be no more than $1$, corresponding to a maximal variance of $\sigma_M^2 \lesssim \frac{8}{81} \xi \big(\frac{k_*}{k_p}\big)^{n_s-1} \big(\frac{k_*}{k_M}\big)^4e^{-(k_*/k_M)^2}$, within which our analysis is restricted \footnote{The non-perturbative regime can be reached in the sound speed resonance (see Fig.~\ref{fig:ContourPlot1}), which is also interesting for PBH formation, but is beyond the scope of the current study.}.

\begin{figure}[h!]
\begin{center}
\includegraphics[width=0.5\textwidth]{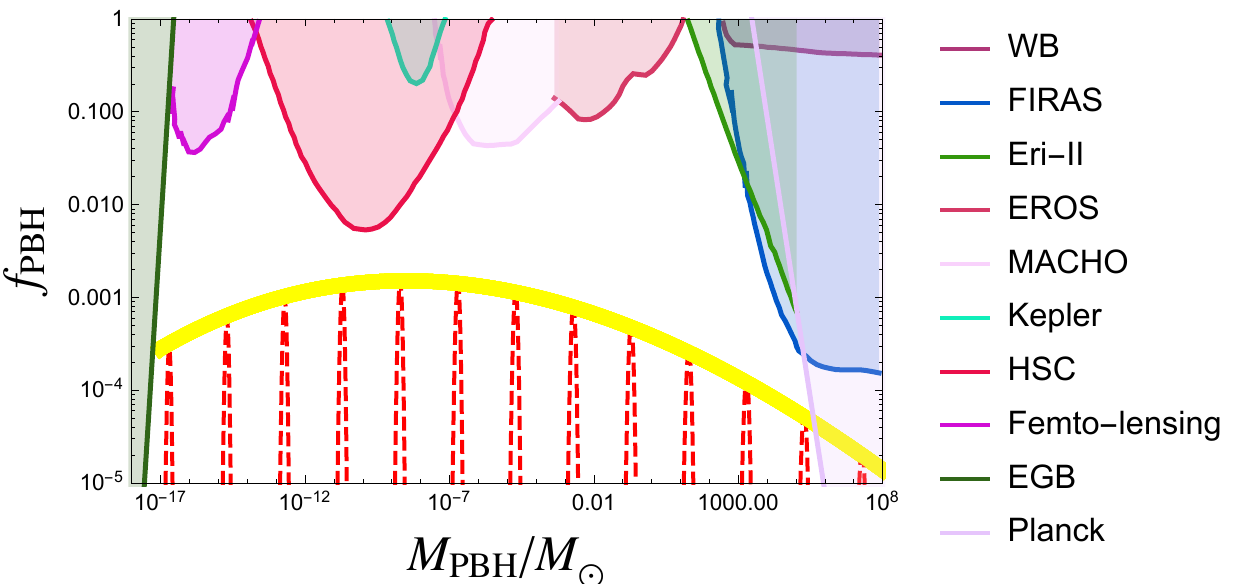}
\end{center}
\caption{
Estimations for the fraction of PBHs against the total DM density, $f_\text{PBH}$, produced by sound speed resonance, for different values of $k_*$. Constraints from a number of astronomical experiments are also shown (see main text for refs.): their observational sensitivities are given by colored shadow areas. We choose $\xi = 0.1$ as well as a group of typical values for the rest parameters: $\gamma=0.2$, $g_{\text{form}} \simeq 100$, $\delta_c=0.37$, $n_s=0.968$.
}
\label{fig:fPBHPlot1}
\end{figure}

PBHs formed by sound speed resonance can account for dark matter in wide parameter ranges and easily satisfy experimental bounds. To see this, we consider the fraction of PBHs against the total dark matter component at present \cite{Sasaki:2018dmp}:
\begin{align} \label{fPBH1}
 & f_\text{PBH}(M) \equiv \frac{\Omega_{\rm PBH}}{\Omega_{\rm DM}} \\
 & = 2.7\times 10^8 \Big(\frac{\gamma}{0.2}\Big)^{1/2} \Big(\frac{g_{*,\text{form}}}{10.75}\Big)^{-1/4}\Big(\frac{M}{M_{\astrosun}}\Big)^{-1/2}\beta(M)~, \nonumber
\end{align}
where $g_{*,\text{form}}$ is the total relativistic degrees of freedom at the PBH formation time.

Numerical estimations of $f_\text{PBH}$ are shown in Fig.~\ref{fig:fPBHPlot1}, for $\gamma=0.2$, $g_{\text{form}} \simeq 100$ \cite{Carr:2009jm} and $\delta_c=0.37$ \cite{Sasaki:2018dmp},  representative of the physics of a typical PBH formation, as well as adopting the Planck result for $n_s=0.968$ \cite{Ade:2015lrj}, and choosing the oscillation amplitude $\xi = 0.1$. We also show current bounds of various astronomical experiments including EGB (extragalactic $\gamma$-ray background), microlensing of Kepler, HSC (Hyper Suprime-Cam), MACHO (massive astrophysical compact halo object), EROS (Expérience pour la Recherche d'Objets Sombres), FIRAS (The Far Infrared Absolute Spectrophotometer) and the Planck satellite \cite{Carr:2016drx}. In Fig.~\ref{fig:fPBHPlot1}, the red dashed lines correspond to the predictions of $f_\text{PBH}$ with different choices for the resonance frequency $k_*$. The PBH mass distribution is given by a narrow peak around $k_*$: this is a distinctive feature of PBHs formed by sound speed resonance from PBHs formed by other processes, for which the mass distribution is usually more spread out. By varying the value of $k_*$, the peaks form a one-parameter family enveloped by a yellow solid curve that mainly depends on the amplitude parameter $\xi$.

Because the PBHs formed by sound speed resonance possess a very narrow mass distribution, no particular tuning of the background is needed to generate PBHs in abundance, consistently with current experimental bounds. From previous discussion, we know that the resonance frequency $k_*$ provides the median of the PBHs mass distribution $M$, while $f_{\rm PBH}$ is mainly determined by the oscillation amplitude $\xi$, and the e-folding numbers $\Delta N$ from $\tau_0$ to the horizon-exit time of $k_*$ mode. Through $f_\text{PBH}$ in Eq. \eqref{fPBH1}, these model parameters can be bounded by various astronomical constraints. In Fig.~\ref{fig:ContourPlot1}, we plot contours for different $\Delta N$, above which the parameter space is excluded by various astronomical constraints. One can see that, even within the scope of the perturbative treatment we followed, the sound speed resonance has a large parameter space, left to be probed by future observations.

\begin{figure}[h!]
\centering
\includegraphics[width=0.3\textheight]{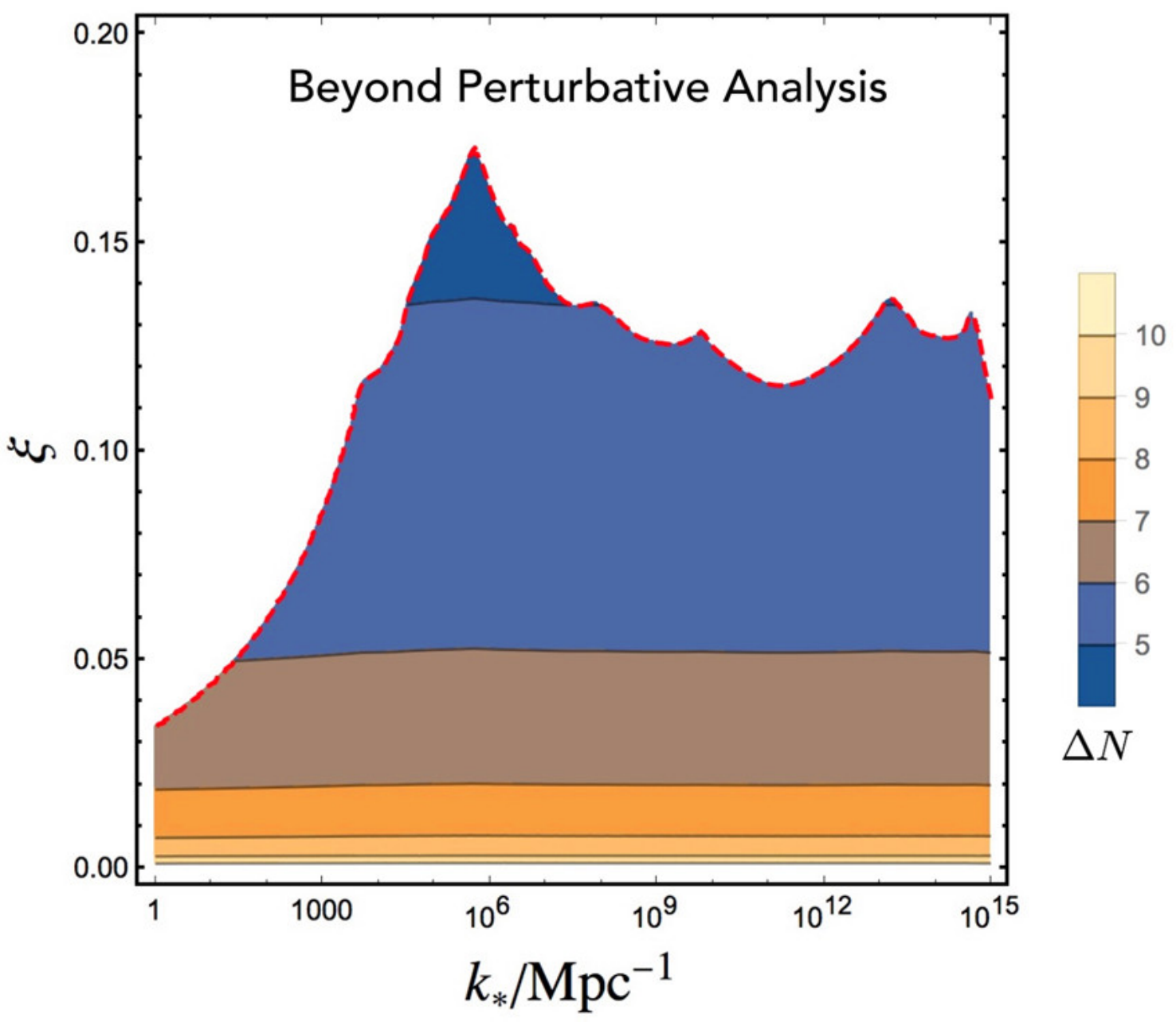}
\caption{
Constraints on the parameter space of sound speed resonance from various astronomical experiments shown in Fig.~\ref{fig:fPBHPlot1}. The white regime is excluded since the enhancement yields $\zeta(k_*) >1$ which breaks the perturbation theory.
}
\label{fig:ContourPlot1}
\end{figure}

%\section{Conclusion}
{\it Conclusions.--}
In this {\it Letter} we proposed a novel mechanism generating PBHs from resonating primordial density perturbations in inflationary cosmology with an oscillatory feature in the sound speed of their propagation.
This scenario may be realized in the context of effective field theory of inflation or by non-canonical models inspired by string theory. Using a parametrization of the oscillating sound speed, our analysis demonstrates that primordial curvature perturbations could be resonantly enhanced in a narrow band of comoving wavenumbers around the oscillation frequency of the sound speed. Consequently, the power spectrum of primordial density fluctuations presents a sharp peak around this resonance frequency, while remains nearly scale-invariant on large scales as predicted by inflationary cosmology. Accordingly, a considerable amount of PBHs can eventually form when the amplified modes re-enter the Hubble radius and may be testable in forthcoming astronomical surveys.

Note that, with this mechanism, enhancement of primordial density fluctuations on specific small scales can be extremely efficient in comparison with other existing mechanisms for PBH formation. {{Another profound property is that, the parametric resonance in our mechanism is a {\it narrow} resonance case and is directly working on curvature perturbation without leading to nonlinear growth of entropy field fluctuations, which can avoid unwanted backreaction effect from a second field perturbation.}} Besides, PBHs generated by sound speed resonance can easily account for DM in current experimental bounds, especially since their mass distribution is very narrow.

We end by highlighting the implications of the proposed mechanism that could initiate future studies from several perspectives. First of all, in this work we mainly study the first peak in the power spectrum, but as discussed, the parametric resonance effect also gives rise to discrete peaks on smaller scales. Although they are not as significant as the first one, it is still possible to have PBH formation on these higher harmonic scales, therefore our model may yield a distinct feature for PBHs mass distribution. Phenomenologically, an important lesson from our study is that, as we started with small oscillations in the sound speed of the propagation of primordial curvature fluctuations in inflation, and ended up with a dramatic production of PBHs, the observational windows on the early universe are no longer limited within the CMB and LSS surveys, but also include other astronomical instruments probing at much smaller scales. On one hand, this motivates theoretical investigations on the possible inflation models from fundamental theories or effective field theories, which could yield oscillating behaviours in the sound speed. Moreover, it is important to further explore how a general time-varying sound speed may affect the evolutions of primordial density fluctuations nonlinearly. On the other hand, in the era of multi-messenger astronomy, PBHs are becoming more and more testable, making for a more and more serious DM candidate, which may inspire designs for future experiments. In particular, detection of GWs produced in black holes merger events could provide great insights on the black holes distribution and their masses.

%\section*{Acknowledgments}
{\it Acknowledgments.--}
We are grateful to
A. Ach{\'u}carro, R. Brandenberger, X. Chen, M. Sasaki and P. Zhang
for stimulating discussions and valuable comments.
This work is supported in part by the National Youth Thousand Talents Program of China, by the NSFC (Nos. 11722327, 11653002), by the CAST Young Elite Scientists Sponsorship (2016QNRC001), and by the Fundamental Research Funds for Central Universities.
DGW is supported by a de Sitter Fellowship of the Netherlands Organization for Scientific Research (NWO).
All numerics were operated on the computer cluster LINDA in the particle cosmology group at USTC.\\

{\it This Letter is dedicated to the memory of the giant Prof. Stephen Hawking, who inspired numerous young people to pursue the dream about the Universe, and beyond.}

\bibliography{cosmo}
\end{document}